\documentclass[onecolumn,
showpacs,preprintnumbers,amsmath, amsfonts]{revtex4-2}
\usepackage{eso-pic}
\usepackage{graphicx}
\usepackage{color}
\usepackage{type1cm}
\usepackage{natbib}

\newsavebox{\longversionbox}
\begin{document}
%\preprint{APS/123-QED}
\title{Infinite Statistics and the Gross Pitaevskii Equation}
% Force line breaks with \\

%\draft
\author{Dewi Yustikasari}
\email{dewiyustikasari96@mail.ugm.ac.id}
\affiliation{Department of Physics, Gadjah Mada University}
\author{Mirza Satriawan}
\email{mirza@ugm.ac.id}
\affiliation{Department of Physics, Gadjah Mada University}
\date{\today}% It is always \today, today,
\begin{abstract}
Abstract: 
We clarify that an ideal gas obeying infinite statistics cannot undergo condensation.  Then we derive the dynamic equation for an identical particle system obeying infinite statistics under external potential and inter-particle interaction.  The derivation utilizes the Hamiltonian written in terms of the number operators and the transition number operators.  At a very low temperature, where one can discard the dynamics of the excited occupation level, the dynamic of an infinite statistics system can be described by the Gross Pitaevskii equation, similar to the Bose-Einstein case. 
\end{abstract}

\maketitle

Even though we have not observed any elementary particle obeying statistics besides the Fermi and the Bose statistics, there are many suggestions that infinite statistics \cite{doplicher1971local, doplicher1974local, govorkov1981parastatistics, greenberg1990example, greenberg1991particles} may indeed be the statistics of certain cosmological objects. Among others are a system of charged extremal black holes \cite{strominger1993black,minic1997infinite}, a system of dark energy quanta in the context of space-time foam cosmology \cite{ng2007holographic, jejjala2007fine, medved2009comment, ng2012holographic} and in the context of M-theory \cite{jejjala2007fine}, a system of dark matter condensate \cite{ebadi2013infinite}, and a system of quanta of modified newton dynamics-dark matter \cite{ho2012dark}.   In infinite statistics all particle permutation symmetry types (representations of the permutation group) are allowed, consequently, not only the ground state but all occupation levels can have a macroscopic number of particles.  This raises a question of whether in infinite statistics one can have a condensation like in the Bose-Einstein condensation, where a macroscopic number of particles resides in the ground state.   In \cite{goodison1994canonical} a critical temperature related to the vanishing of the chemical potential in infinite statistics is given, but vanishing chemical potential does not always mean a condensation in the ground state has take place.  In \cite{ebadi2013infinite} the critical temperature for condensation is derived using the counting function of Medvedev ambiguous statistics \cite{medvedev1997properties}, but Medvedev's counting function is not the counting function for infinite statistics \cite{meljanac1999infinite}.  

The correct dynamics of a real particle system obeying infinite statistics may include external potential and inter-particle interaction.  For such a system that obeys Bose-Einstein statistics, the dynamics can be described by the Gross Pitaevskii equation \cite{gross1961structure,pitaevskii1961vortex,dalfovo1999theory}.   While for infinite statistics similar equation is not yet known.  In this paper, after clarifying that condensation cannot take place in a system of ideal gas obeying infinite statistics, we continue to derive the dynamic equation for an infinite statistics system under external potential and inter-particle interactions.  We show that if one only considers the ground state, the dynamics of an infinite statistics system can be described by the Gross Pitaevskii equation similar to the Bose-Einstein condensation.      

Consider $N$ ideal gas particles obeying infinite statistics.  The partition function for this system can be known once the states in the system are determined. The partition function is given as usual by
\begin{equation}
    Z_N = \sum_{\rm allowed\ states} e^{-\beta H}
\end{equation}
where $\beta = 1/kT$ and $H$ is the Hamiltonian.  Since all symmetry types are allowed in infinite statistics, for a system with $n_i$'s particles in the $i$-th occupancy level, the number of orthogonal states is given by~\cite{greenberg1990example}
\begin{equation}
    g(\{n_i\}) = \frac{N!}{\prod_i n_i!}.
\end{equation}
Thus, the partition function is given by
\begin{equation}
    Z_N = \sum_{n_1+n_2\cdots = N} \frac{N!}{\prod_i n_i!} \prod_i e^{-\beta E_i n_i} = (1+e^{-\beta E_1}+ e^{-\beta E_2}  +\cdots)^N
\end{equation}
where $E_i$ is the energy of the i-th occupation level.  For an open system, where the number of particles is not fixed, the thermodynamics can be described from the grand canonical partition function.  Knowing the above partition function, the grand canonical partition function for an ideal gas system obeying infinite statistics is given by
\begin{equation}
\mathcal{Z} = \frac{1}{1-(x_1+x_2+x_3+\dots)}
\end{equation}
where $x_i = e^{\beta(\mu - E_i)}$.  The average number of particles in each occupation level and the average total number of particles are given respectively by
\begin{equation}
N_k = -\ \frac{1}{\beta}\frac{\partial}{\partial E_k}\log Z = \frac{x_k}{\left(1 - \sum_{i=1} x_i \right)}; \quad N =   \frac{\sum_{k=1} x_k}{\left(1 - \sum_{i=1} x_i \right)} 
\end{equation}
Moving to the continuum limit, the sum over energy levels should be replaced by an integral over the invariant phase space.  After separating the contribution from the ground state, we have 
\begin{equation}
     \sum_{i=1} ze^{-\beta E_i} \rightarrow  a + \frac{2 \pi^{3/2}V}{h^3 \Gamma(3/2)} \int_0^\infty 
\ z e^{-\beta E}\ p^{2} dp = a(1 + VL)
\end{equation}
where $a = z e^{-\beta mc^2}$, $E = \sqrt{m^2c^4 + p^2}$, $m$ is the mass of the particle, and following Beckmann et.al we have introduce ~\cite{beckmann1979bose}
\begin{equation}\label{a:MBcase}
L \equiv  \ \beta \frac{2c}{\hbar^3}\left(\frac{m}{2\pi\beta}\right)^{2}e^u K_{2}(u).
\end{equation}
with $u \equiv \beta mc^2$ and $K_2(u)$ is the second order modified Bessel functions of the second kind.  In the non-relativistic limit ($u \rightarrow \infty$), we have $L \rightarrow \lambda_T^{-3}$, where $\lambda_T = \sqrt{2 \pi \hbar^2\beta / m}$ is the thermal wavelength.   In the ultra-relativistic limit ($u \rightarrow 0$), we have $L \rightarrow 1/\pi^2(c\beta \hbar )^3$.   The average total energy and the total number of particles in the continuum limit are given respectively by
\begin{equation}
\begin{split}
    U &= \frac{2 \pi^{3/2}V}{h^3 \Gamma(3/2)}\frac{1}{1-a(1 + VL)} \int_0^\infty 
\ z E e^{-\beta E}\ p^{2} dp  = 
\left(3NkT + Nmc^2 \frac{K_{1}(u)}{K_{2}(u)}  \right) \\
    N(T,V,\mu)&= N_0 + N_e ; \quad N_0 = \frac{a}{\left(1 - a(1 + VL)\right)}; \quad N_e = \frac{aVL}{\left(1 - a(1 + VL)\right)} 
\end{split}
\end{equation}
In the thermodynamics limit (large $N$), we have
\begin{equation}
    a(1+VL) = \frac{N}{N+1} \simeq 1.
\end{equation}
Since the value of $VL$ has to be positive, we have a constraint that $a< 1$.  In the limit when $a \rightarrow 1$, the value $VL \rightarrow 0$, and this take place when $T \rightarrow 0$ for all occupation level.   The ratio between the excited states and the ground state occupation number is given by $N_e/N_0 = VL$.  The particles will dominantly be in the ground states when $VL \rightarrow 0$, that is when $T\rightarrow 0$, i.e. there is no critical temperature and there is no condensation.  In the thermodynamics limit all occupation levels will always have a large number of particles.  At high temperatures, most particles will be at the excited level.  By decreasing the temperature particles will tend to occupy a lower occupation level, and at zero temperature, all particles will be in the ground state.

The dynamics of particles obeying infinite statistics can be derived using the Heisenberg equation of motion for the field operator or the creation annihilation operator.  The creation-annihilation operator algebra realization for the infinite statistics is already known, formulated a long time ago by Greenberg \cite{greenberg1990example}
\begin{equation}
a_i a_j^\dagger - q a_j^\dagger a_i = \delta_{ij}
\end{equation}
where $a_i$ and $a_j^\dagger$ are the annihilation and creation operator respectively, and $q$ is a real parameter whose value is $-1 < q < 1$.  But it is not easy to derive the equation of motion using the infinite statistics creation annihilation operator algebra above.  Instead, it turns out easier to use the number operator and the transition number operator in the Heisenberg equation of motion.  The particle transition number operator of the infinite statistics, for the $q=0$ case, is given by \cite{greenberg1990example} 
\begin{equation}\label{transition}
n_{ij} = a_i^\dagger a_j +\sum_k a_k^\dagger a_i^\dagger a_j a_k + \sum_{k_1,k_2} a_{k_1}^\dagger a_{k_2}^\dagger a_i^\dagger a_j a_{k_2}a_{k_1}+\cdots +\sum_{k_1,k_2,\dots,k_s}a_{k_1}^\dagger a_{k_2}^\dagger\dots a_{k_s}^\dagger a_i^\dagger a_j a_{k_s}\dots a_{k_2}a_{k_1}+\cdots.
\end{equation}
The case when $i=j$ above will give the number operator.  Note that the transition number operator in \eqref{transition} obey the usual commutation relation with the annihilation operator $[n_{ij},a_k] = -\delta_{ik} a_j$.  

We start the derivation from the description of the Hamiltonian in terms of the particle number operators and the particle transition number operators.   We assume that we only deal with particles with low kinetic energy, thus we use the non-relativistic formulation for the kinetic energy.  The Hamiltonian is given by
\begin{equation}\label{Hamiltonian}
\begin{split}
H &= \int \frac{d^3 \mathbf{k}}{(2\pi)^3} \frac{\hbar^2 k}{2m}  n(\mathbf{k},t) + \int \frac{d^3 \mathbf{k}}{(2\pi)^3} \int \frac{d^3 \mathbf{k}'}{(2\pi)^3}  V_{ext}(\mathbf{k}-\mathbf{k}')n(\mathbf{k},\mathbf{k}',t)\\
&+ \frac{1}{2} \int \frac{d^3 \mathbf{k}}{(2\pi)^3} \int \frac{d^3 \mathbf{k}'}{(2\pi)^3} \int \frac{d^3 \mathbf{k}''}{(2\pi)^3} \int \frac{d^3 \mathbf{k}'''}{(2\pi)^3}   n(\mathbf{k}'',\mathbf{k},t)V(|\mathbf{k}-\mathbf{k}''|)n(\mathbf{k}''',\mathbf{k}',t) (2\pi)^3\delta^{3}(\mathbf{k} + \mathbf{k}' - \mathbf{k}'' - \mathbf{k}''')
\end{split}
\end{equation}
where $n(\mathbf{k},t)$ is the particle number operator with particle's momentum $\mathbf{k}$, $n(\mathbf{k},\mathbf{k}',t)$ is the particle transition number operator from momentum  $\mathbf{k}'$  into $\mathbf{k}$. The first term in the right hand side of \eqref{Hamiltonian} is the kinetic energy terms, the second term is the interaction energy with an external potential $V_{ext}(\mathbf{k} -\mathbf{k}')$, and the third term is a two-particle interaction energy, with initial particle momenta are $\mathbf{k}$ and $\mathbf{k}'$ and final momenta are $\mathbf{k''}$ and $\mathbf{k}'''$.  The $V(|\mathbf{k}-\mathbf{k}''|)$ is the inter-particle interaction potential, where we have assumed that the interaction is through a t-channel interaction, and lastly, the delta Dirac function is to impose the total momentum conservation.

The Heisenberg equation of motion for the annihilation operator is given by
\begin{equation}\label{heisenberginfinite}
\begin{split}
 i\hbar \frac{\partial}{\partial t} a(\mathbf{k},t) & = \left[a(\mathbf{k},t), H\right] \\
  i\hbar \frac{\partial}{\partial t} a(\mathbf{k},t) & =\frac{\hbar^2 \mathbf{k}^2}{2m} a(\mathbf{k},t) + \int \frac{d^3 \mathbf{k}'}{(2\pi)^3} V_{\rm ext}(\mathbf{k}-\mathbf{k}') a(\mathbf{k}',t) \\
& +\int \frac{d^3 \mathbf{k}'}{(2\pi)^3} \int \frac{d^3 \mathbf{k}''}{(2\pi)^3}  n(\mathbf{k}''+ \mathbf{k}'-\mathbf{k},\mathbf{k}',t) a(\mathbf{k}'',t) V(|\mathbf{k}-\mathbf{k}''|) + 
\frac{1}{2}\int \frac{d^3 \mathbf{k}'}{(2\pi)^3} a(\mathbf{k},t)V(|\mathbf{k} - \mathbf{k}'|)\\
\end{split}
\end{equation}
Switching into position basis on Fourier transformation, we have
\begin{equation} \label{Heisenberginfinite1}
i \hbar \frac{\partial}{\partial t} \Psi (\mathbf{r},t) = \left(-\frac{\hbar^2}{2m} \nabla^2 + V(\mathbf{r})_{\rm ext} +\frac{V(0)}{2}  \right) \Psi (\mathbf{r},t) + \int d^3 \mathbf{r}'  n(\mathbf{r}',t) \Psi (\mathbf{r},t)V(\mathbf{r}'-\mathbf{r})  
\end{equation}
where $\Psi(\mathbf{r},t)$ is the field operator, and $n(\mathbf{r}',t)$ is the particle number operator in the position basis.   The $V(0)$ term above is just a constant in the potential energy and can be discarded without affecting the equation of motion.  
%In the low energy case, the statistics of the particles affect the dynamics from the inter-particle interaction, i.e. from the particle transition number operator formulation in terms of the %field operators.  For the Bose-Einstein and Fermi-Dirac statistics the transition operator in terms of the field operators is given by $n(\mathbf{r}',\mathbf{r},t) = \Psi^\dagger %(\mathbf{r}',t) \Psi (\mathbf{r},t)$, with the field operators obey the commutation and the anti commutation relations respectively.   
We can expand the field operator around its value at the ground state as
\begin{equation}\label{condensatfield}
\Psi(\mathbf{r},t) = \psi(\mathbf{r},t) + \Psi'(\mathbf{r},t) 
\end{equation}
where $\psi(\mathbf{r},t) = \langle \Psi(\mathbf{r},t) \rangle$ is the value of the field at the ground state, i.e. the ground state classical field, while $\Psi'(\mathbf{r},t)$ is the quantum excitation field.  At very low temperatures, a large amount of the particles are in the ground state, therefore we can neglect the contribution from the excited states and assume the system is described only by its ground state field.  Using Eq.\eqref{condensatfield} and taking the ground state expectation values of Eq. \eqref{Heisenberginfinite1}  we have 
\begin{equation} \label{Heisenberginfinite2}i \hbar \frac{\partial}{\partial t} \psi (\mathbf{r},t) = \left(-\frac{\hbar^2}{2m} \nabla^2 + V(\mathbf{r})_{\rm ext} \right) \psi (\mathbf{r},t) + \int d^3 \mathbf{r}'    n(\mathbf{r}',t)\psi (\mathbf{r},t) V(\mathbf{r}'-\mathbf{r}) 
\end{equation}

Taking the Fourier transform of \eqref{transition}, and then using \eqref{condensatfield} neglecting the excited states, we can write the particle number for infinite statistics as
\begin{equation}\label{number3}
\begin{split}
   n(\mathbf{r}',t)  
   & = \psi^\dagger (\mathbf{r}',t) \psi (\mathbf{r}',t) \left(1+\int d^3 \tilde{\mathbf{r}} |\psi(\tilde{\mathbf{r}},t)|^2 + \left(\int d^3 \tilde{\mathbf{r}} |\psi(\tilde{\mathbf{r}},t)|^2 \right)^2 + \cdots \right) \\
   &= \frac{\psi^\dagger (\mathbf{r}',t) \psi (\mathbf{r}',t)}{1- \int d^3 \tilde{\mathbf{r}} |\psi(\tilde{\mathbf{r}},t)|^2}\end{split}.
\end{equation}
Taking the integral of the above over the whole space, we have the average total number of particles 
\begin{equation}
    N = \frac{\int d^3 \mathbf{r} |\psi (\mathbf{r},t)|^2}{1- \int d^3 \mathbf{r} |\psi(\mathbf{r},t)|^2}.
\end{equation}
Thus, in the thermodynamics limit, we have
\begin{equation}
    \int d^3 \mathbf{r}|\psi (\mathbf{r},t)|^2 = \frac{N}{N+1}\simeq 1
\end{equation}
and we can write the particle number as
\begin{equation}\label{number4}
     n(\mathbf{r}',t)  \simeq N |\psi(\mathbf{r}',t)|^2
\end{equation}
Inserting Eq.\eqref{number4} into \eqref{Heisenberginfinite2}, we have 
\begin{equation}\label{dynamic}
i \hbar \frac{\partial}{\partial t} \psi (\mathbf{r},t) = \left(-\frac{\hbar^2}{2m} \nabla^2 + V(\mathbf{r})_{\rm ext} \right) \psi (\mathbf{r},t) + \int d^3 \mathbf{r}'  
N |\psi(\mathbf{r}',t)|^2 \psi (\mathbf{r},t) V(\mathbf{r}'-\mathbf{r})  
\end{equation}
If we renormalize the wave function condensate as
\begin{equation}
    \psi (\mathbf{r},t)  \rightarrow \frac{1}{\sqrt{N}}\psi (\mathbf{r},t) 
\end{equation}
we will get
\begin{equation}\label{dynamic2}
i \hbar \frac{\partial}{\partial t} \psi (\mathbf{r},t) = \left(-\frac{\hbar^2}{2m} \nabla^2 + V(\mathbf{r})_{\rm ext} \right) \psi (\mathbf{r},t) + \int d^3 \mathbf{r}'  |\psi(\mathbf{r}',t)|^2 \psi (\mathbf{r},t) V(\mathbf{r}'-\mathbf{r})  
\end{equation}
with now,
\begin{equation}
    \int d^3 \mathbf{r}|\psi (\mathbf{r},t)|^2 = N
\end{equation}
In particular for interparticle interaction in the form of four point interaction, $V(\mathbf{r}'-\mathbf{r}) = \lambda \delta^3(\mathbf{r}'-\mathbf{r})$ we have
\begin{equation}\label{dynamic3}
i \hbar \frac{\partial}{\partial t} \psi (\mathbf{r},t) = \left(-\frac{\hbar^2}{2m} \nabla^2 + V(\mathbf{r})_{\rm ext} \right) \psi (\mathbf{r},t) + \lambda |\psi(\mathbf{r},t)|^2 \psi (\mathbf{r},t), 
\end{equation}
this is just the Gross Pitaevksii equation.  In conclusion, the dynamics of identical particles system obeying infinite statistics at low energy can be described by the usual Gross Pitaevskii equation.  This last result is expected since a system with one occupation state (the ground state) obeying infinite statistics should not be different from a Bose-Einstein system.

\bibliographystyle{unsrt}
\bibliography{main.bib}

\end{document}